\documentclass{aastex61}

\usepackage{CJK}
\usepackage{amssymb}
\usepackage{multirow}
\usepackage{morefloats}
\usepackage{amsmath}
\usepackage{bigstrut}
\usepackage{booktabs}
\usepackage{natbib}
\usepackage{float}
\usepackage{graphicx,epsfig,fancyhdr,epsf,txfonts,epstopdf}
\usepackage{latexsym,bbm}
\usepackage{lineno}
\usepackage{color}
\usepackage{ulem}
\usepackage{threeparttable}
\usepackage{longtable}

\newcommand{\ccc}{\emph{c}}
\newcommand{\Tb}{$T_0$ }
\newcommand{\vc}{$v_e$ }

\newcommand{\gammamax}{$\gamma_{max}$ }
\newcommand{\omegarat}{$\omega_{pe} / \Omega_{ce}$ }

\received{***}
\revised{***}
\accepted{***}

\shortauthors{Li et al.}
\shorttitle{Warm-plasma effect on Z-mode excitation}
\begin{document}

\title{Effect of the temperature of background plasma and the energy of energetic electrons on Z-mode excitation}

\correspondingauthor{Yao Chen}
\email{yaochen@sdu.edu.cn}

\author{Chuanyang Li}\thanks{E-mail: licy001@163.com}
\affil{Institute of Space Sciences and Shandong Provincial Key Laboratory of Optical Astronomy and Solar-Terrestrial Environment \\
Shandong University, Weihai, Shandong 264209, China\\}

\author{Yao Chen}\thanks{E-mail: yaochen@sdu.edu.cn}
\affiliation{Institute of Space Sciences and Shandong Provincial Key Laboratory of Optical Astronomy and Solar-Terrestrial Environment \\
Shandong University, Weihai, Shandong 264209, China\\}

\author{Xiangliang Kong}
\affiliation{Institute of Space Sciences and Shandong Provincial Key Laboratory of Optical Astronomy and Solar-Terrestrial Environment \\
Shandong University, Weihai, Shandong 264209, China\\}

\author{M. Hosseinpour}
\affiliation{Institute of Space Sciences and Shandong Provincial Key Laboratory of Optical Astronomy and Solar-Terrestrial Environment \\
Shandong University, Weihai, Shandong 264209, China\\}
\affiliation{Faculty of Physics, University of Tabriz, Tabriz, Iran}

\author{Bing Wang}
\affiliation{Institute of Space Sciences and Shandong Provincial Key Laboratory of Optical Astronomy and Solar-Terrestrial Environment \\
Shandong University, Weihai, Shandong 264209, China\\}


\begin{abstract}
It has been suggested that the Z-mode instability driven by energetic electrons with a loss-cone type velocity distribution is one candidate process behind the continuum and zebra pattern of solar type-IV radio bursts. Both the temperature of background plasma ($T_0$) and the energy of energetic electrons ($v_e$) are considered to be important to the variation of the maximum growth rate ($\gamma_{max}$). Here we present a detailed parameter study on the effect of $T_0$ and $v_e$, within a regime of the frequency ratio ($10 \leq \frac{\omega_{pe}}{\Omega_{ce}} \leq 30$). In addition to $\gamma_{max}$, we also analyze the effect on the corresponding wave frequency ($\omega^r_{max}$) and propagation angle ($\theta_{max}$). We find that (1) $\gamma_{max}$ in-general decreases with increasing $v_e$, while its variation with $T_0$ is more complex depending on the exact value of $v_e$; (2) with increasing $T_0$ and $v_e$, $\omega^r_{max}$ presents step-wise profiles with jumps separated by gradual or very-weak variations, and due to the warm-plasma effect on the wave dispersion relation $\omega^r_{max}$ can vary within the hybrid band (the harmonic band containing the upper hybrid frequency) and the band higher; (3) the propagation is either perpendicular or quasi-perpendicular, and $\theta_{max}$ presents variations in line with those of $\omega^r_{max}$, as constrained by the resonance condition. We also examine the profiles of $\gamma_{max}$ with $\frac{\omega_{pe}}{\Omega_{ce}}$ for different combinations of $T_0$ and $v_e$ to clarify some earlier calculations which show inconsistent results.
\end{abstract}


\section{Introduction}
Recent studies on moving type-IV solar radio bursts (t-IVms, slowly-drifting wide band continuum observed at metric-decimetric wavelengths) reveal that the t-IVm sources are associated with an eruptive high-temperature dense structure (\citealp{Vasanth16,Vasanth19}). This is possible since the events of study are recorded at metric wavelengths by both the \emph{Nan\c{c}ay Radioheliograh} (NRH: \citealp{Kerdraon97}) and at Extreme Ultraviolet (EUV) by the Atmospheric Imaging Assembly on board the \emph{Solar Dynamics Observatory} (AIA/SDO: \citealp{Lemen12,Pesnell12}). Further differential emission analysis of AIA data shows that the source temperature is around several MK and the density is at the level of $10^8$ cm$^{-3}$ at a heliocentric distance of $\sim$1.2--1.5 Solar Radii (R$_\odot$) at frequencies around 200-300 MHz (\citealp{Vasanth19}). At this height of the solar atmosphere, the magnetic field strength is in general around or less than several Gauss (see, e.g., \citealp{Dulk78, Cho07, Ramesh10, Chen11, Feng11}). Thus, the metric t-IVm bursts are generated within a plasma regime with the plasma-electron-cyclotron frequency ratio ($\frac{\omega_{pe}}{\Omega_{ce}}$) much larger than unity (mostly larger than 10). Based on the observations and some related earlier theoretical studies (e.g., \citealp{Winglee86,Benacek17}), \cite{Vasanth16, Vasanth19} suggested that the t-IV continuum belongs to coherent plasma emission generated by energetic electrons trapped within the eruptive magnetic structure.

Energetic electrons trapped by a magnetic structure can develop a loss-cone type distribution with an inversion of population along the perpendicular direction in the velocity space, i.e., $\frac{\partial f}{\partial v_\perp} > 0$, where $f$ represents the velocity distribution function of energetic electrons. They can drive kinetic instabilities and excite plasma waves (e.g., \citealp{Freund76, Freund77, Wu79, Wu85, Winglee86}). In the parameter regime of $\frac{\omega_{pe}}{\Omega_{ce}} \gg 1$, such distribution can result in the Z-mode instability and excites enhanced Z-mode waves, which are the slow branch of the extraordinary (X) mode and corresponding to obliquely (or perpendicularly) propagating Langmuir waves. Under certain conditions, such as propagating in inhomogeneous magnetic field and nonuniform plasmas, Z-mode waves may transform into escaping electromagnetic mode and be observed as radio bursts such as t-IV bursts (e.g., \citealp{Winglee86}).

Many earlier studies have applied the Z-mode instability to explain the origin of the intriguing embedding zebra structure of t-IV bursts (e.g., \citealp{Winglee86, Yasnov04, Zlotnik13, Benacek17}). Zebras refer to the numerous emission stripes that are almost parallel to each other superposed on the t-IV continuum, as manifested on the solar radio dynamic spectra (\citealp{Kundu65,Slottje72,Kruger79,Chernov01,Chernov10,Chernov12,Tan14}). Note that the Z-mode instability is also called as the double-plasma resonance (DPR) in many references (\citealp{Yasnov04,Benacek17,Benacek18}), since for cold plasmas the instability reaches the maximum growth rate when the upper hybrid frequency ($\omega_{UH}=\sqrt{\omega_{pe}^2 + \Omega_{ce}^2}$) equals a harmonic of $\Omega_{ce}$. It is found that the most important parameter relevant to solar radio bursts is the frequency ratio ($\frac{\omega_{pe}}{\Omega_{ce}}$) which strongly modulates the values of the maximum of the Z-mode growth rate ($\gamma_{max}$). With increasing $\frac{\omega_{pe}}{\Omega_{ce}}$, the profile of $\gamma_{max}$ manifests peaks at frequencies close to harmonics of $\Omega_{ce}$, i.e., $\gamma_{max}$ reaches maximum when Z-mode frequency is close to $n\Omega_{ce}$ where $n$ is an integer. Note that the exact mechanism(s) accounting for the t-IVm continuum and zebras are yet to be determined. For the continuum, both incoherent gyro-synchrotron and coherent plasma emission have been suggested, while more scenarios exist for zebras, such as the Bernstain wave mode, whistler wave mode, the DPR or the Z-mode instability (see the review in \citealp{Chernov10} and the lastest statistical study by \citealp{Tan14}). In this study, only the last possibility, i.e., the mechanism involving the Z-mode instability, will be explored.

\cite{Winglee86} is one of the first to investigate the effect of $\frac{\omega_{pe}}{\Omega_{ce}}$ on $\gamma_{max}$ of Z-mode. They proposed that the Z-mode instability driven by energetic electrons with loss-cone type distribution might be able to explain both features. The zebra pattern is due to the peak of the growth rate at certain frequencies that are separated by electron gyro-frequency. Between these peaks, the Z-mode can still be excited though at a lower growth rate. The continuum is explained with inhomogeneity of the magnetic field and density within a large-scale source region. Continuous change of background parameters can result in a continuous variation of plasma characteristic frequencies, leading to the Z-mode growth within the corresponding frequency range. According to \cite{Winglee86}, this may serve as a unified scenario for both type-IV continuum and zebras. Using the standard loss-cone distribution and the Dory, Guest, and Harris (DGH, \citealp{Dory65}) distribution for energetic electrons as inputs, they concluded that the latter is likely more relevant to t-IV bursts with significant zebra structures.

Following \cite{Winglee86}, with the DGH distribution \cite{Yasnov04} and \cite{Benacek17} further investigated the effect of background plasma temperature and energy of energetic electrons on the $\frac{\omega_{pe}}{\Omega_{ce}}$ dependence of Z-mode growth rate. Yet, the two studies drew inconsistent conclusions. \cite{Yasnov04} concluded that the zebra pattern with more significant peaks of $\gamma_{max}$ forms for higher energy of energetic electrons, and $\gamma_{max}$ increases with increasing $T_0$ (from 2 MK to 20 MK), while \cite{Benacek17} concluded that the profile of $\gamma_{max}$ versus $\frac{\omega_{pe}}{\Omega_{ce}}$ expresses distinct peaks with relatively low energy of energetic electrons and does not change significantly with $T_0$. Thus, a clarification is required for a consistent understanding of the variation of the growth rate with $\frac{\omega_{pe}}{\Omega_{ce}}$.

In addition, most earlier studies focused on the variation of the growth rate with the plasma-cyclotron frequency ratio, and did not pay much attention to other critical parameters such as the propagation angles ($\theta_{max}$) and the frequencies ($\omega^r_{max}$) at which the instability develops at the maximum growth rate. \cite{Lee13} conducted a parameter study on the electron cyclotron maser instability which directly excites the escaping X and ordinary (O) modes (\citealp{Wu79}) and checked the variation of both parameters for the regime of $\frac{\omega_{pe}}{\Omega_{ce}} \leq 5$. \cite{Yi13} extends the study to investigate the variations of the two parameters of Z-mode instability. Both studies are based on cold plasma approximation. The latter concluded that for $\frac{\omega_{pe}}{\Omega_{ce}} \leq 6$ two Z-mode bands appear for a specific set of parameters with one very narrow band and one relatively wide band. The narrow band is associated with a very short wavelength and may suffer from strong thermal cyclotron damping effect. Thus, it may not survive in warm plasmas, and the wide band may be more relevant to the radio burst. However, the growth rate of the narrow band is often larger than its wide band counterpart, and may be mistakenly picked out when the wave spectrum is not carefully analyzed. This, together with the observational indications from t-IVms (\citealp{Vasanth16,Vasanth19}), points to the significance of including warm-plasma effect when calculating the Z-mode growth rate.

Only a few studies considered such effect on Z-mode growth \citep{Winglee86,Yasnov04,Benacek17}, with various combinations of the background plasma temperature (represented by $T_0$ or the corresponding thermal speed $v_0=\sqrt{\frac{k_BT_0}{m}}$) and energy of energetic electrons (represented by $\emph{v}_e$) of the DGH distribution. Unfortunately, only a few discrete values of the two parameters have been taken into account. For example, \cite{Benacek17} investigated the Z-mode instability for $v_0$ = 0, 0.009, and 0.018 c, and $v_e$ = 0.1, 0.2, and 0.3 c, while \cite{Yasnov04} only considered two different values of $T_0$ (2 and 20 MK). This means that present parameter studies on the effect of $T_0$ and $v_e$ are rather incomplete and more detailed parameter study may be necessary. Indeed, according to the calculations presented below, such study leads to important novel results that have not been reported.

In summary of this section, a detailed parameter study on the effect of the temperature of background plasma and the energy of energetic electrons is required to better understand the Z-mode instability driven by trapped electrons.
This serves as the major motivation of the present study. The following section introduces basic assumptions, the wave dispersion relation (see also the Appendix), and parameters used in the calculations, and in Section 3 results of the parameter study are presented. A summary and discussion are given in the last section.

\section{Basic assumptions, dispersion relation, and parameters}
The present study is based on the general kinetic dispersion relation for small-amplitude waves propagating in uniform magnetized warm plasmas (see, e.g., \citealp{Baldwin69, Wu85}), which is a linear wave solution to the collisionless Vlasov-Maxwell system. The plasma consists of two components of electrons, one is the background warm plasma with the Maxwellian distribution ($f_0$), the other is the energetic electrons with the DGH distribution ($f_e$), given by
\begin{equation}
 f(u_{\perp},u_{\parallel})=\frac{n_e}{n_0}f_e+\left(1-\frac{n_e}{n_0}\right)f_0,
\end{equation}
\begin{equation}
 f_0=\frac{1}{(2\pi)^{3/2}v_0^{3}}\exp\left(-\frac{u^{2}}{2v_0^{2}}\right),
\end{equation}
\begin{equation}
 f_e=\frac{u_\perp^{2j}}{2^j(2\pi)^{3/2}v_e^{3+2j}j!}\exp\left(-\frac{u_\perp^{2}+u_\parallel^{2}}{2v_e^{2}}\right),
\end{equation}
where $f$ is the total electron distribution function, $n_0$ and $n_e$ are number density of thermal and energetic electrons, respectively. \emph{j} is the order of the DGH distribution function ($f_e$) and is set to be 1, $v_e$ is the mean velocity of energetic electrons, $u_\perp \ (u_\parallel)$ represents the averaged perpendicular (parallel) momentum per unit mass of electrons. The ions are assumed to be static since only modes with frequencies much higher than ion characteristic frequencies are considered, thus only electrons contribute to the general dispersion relation.

In Figure 1, we demonstrate the total electron distribution ($f$) with white contours, superposed by maps of $\lg(\partial f/\partial u_\perp +1)$ which represents the $u_\perp$ gradient of $f$. The corresponding temperatures of panels a, b and c are $T_0$ = 0, 2, and 4 MK respectively, and the related electron velocity is fixed at 0.2 c. With increasing $T_0$, $f_0$ occupies a larger area of the velocity space, and gets closer to the phase space occupied by energetic electrons, as expected. This of course affects the growth rate of Z-mode instability which is determined by the integral (see Equation A.5) along the resonance curve as defined in the velocity space by the resonance condition
\begin{equation}
\gamma_L\omega_r-n\Omega_{ce}-k_\parallel u_\parallel=0,
\end{equation}
where $\gamma_L$ is the Lorentz factor, $k_\parallel$ is the parallel wave number, and $\Omega_{ce}$ is the electron cyclotron frequency.

We assume that $n_e \ll n_0$. Then, the wave modes are determined by thermal electrons while the instability is energized by energetic electrons. This allows us to utilize fluid equations of warm plasmas to derive the dispersion relation of Z-mode from which the real part of the wave frequency ($\omega_r=\omega_r(\vec k)$) is deduced, where $\vec k \ (= k \cos\theta \hat{e}_z + k \sin\theta \hat{e}_x)$ represents the wave vector, and $\theta$ is the angle between the background magnetic field $\vec {B}_0 \ (= B_0 \hat{e}_z)$ and $\vec k$. The solution of the growth rate (i.e., the imaginary part of the wave frequency, $\gamma$) can be greatly simplified with this assumption (see Equation A.4). Please check the Appendix for details of $\gamma$, the general kinetic, and fluid dispersion relations.

Both dispersion relations are solved numerically. Major parameters are $T_0$, $v_e$, and $\frac{\omega_{pe}}{\Omega_{ce}}$. The density ratio $\frac{n_e}{n_0}$ is included in the growth rate, therefore its exact value is not important as long as it remains small enough. For $T_0$, considering the t-IVm observations introduced earlier and usual warm-plasma parameters of the solar corona, we vary $T_0$ in a range of [0, 8] MK or $\sim$ [0, 1] keV; for $v_e$ we vary it in a range of [0.15, 0.4] c or $\sim$ [5, 50] keV, where c is the speed of light. Thus, the weakly-relativistic approximation can be applied.

Regarding the range of $\frac{\omega_{pe}}{\Omega_{ce}}$, as mentioned in the introduction this parameter can be much larger than unity for plasmas within the t-IVm sources, we therefore set its range to be [10, 30]. According to observations on t-IV bursts with zebra patterns, the number of stripes and the corresponding estimated harmonics are often larger than 10, sometimes reaching 30 (e.g., \citealp{Kuijpers75, Aurass03, Zlotnik03, Chernov05, Kuznetsov07}). This means the adopted range of $\frac{\omega_{pe}}{\Omega_{ce}}$ is relevant to the study of t-IV radio bursts.

Another reason of using this range of $\frac{\omega_{pe}}{\Omega_{ce}}$ stems from the limitation of the fluid dispersion relation of Z-mode. For cold plasmas, the frequency of Z-mode is determined by the upper hybrid frequency ($\omega_r=\omega_{UH}$). For warm plasmas, its kinetic dispersion relation is greatly affected by the cyclotron resonance effect, being split into many branches (usually called electron Bernstein modes). The branch within the hybrid band (i.e., the band contains $\omega_{UH}$, given by $(s-1) \Omega_{ce}\leq \omega_r \leq s \Omega_{ce}$, $s$ is a positive integer) with normal dispersion corresponds to the Z-mode, the part with abnormal dispersion corresponds to the electron cyclotron mode. Only the normal-dispersion part is of interest here. Dispersion relations given by the fluid equations and the plasma kinetic theory for Z-mode in warm plasmas of  Maxwellian distribution are plotted in Figure 2, for $\frac{\omega_{pe}}{\Omega_{ce}}=$ 5, 10, 15, and 20. It can be seen that for $\frac{\omega_{pe}}{\Omega_{ce}}=$ 5, the fluid dispersion relation deviates away from the kinetic one at frequencies higher than $5.7 \ \Omega_{ce}$, while for larger values ($\geq$ 10) the fluid dispersion relation represents a good approximation to the kinetic one, at least for the normal dispersion parts within the hybrid band and one band higher (i.e., the band of $s \Omega_{ce}\leq \omega_r \leq (s+1) \Omega_{ce}$, see also \Citealp{Zlotnik03}). We therefore use the fluid dispersion relation of Z-mode and limit our discussion to the regime of $10 \leq \frac{\omega_{pe}}{\Omega_{ce}} \leq 30$. As seen from our results, all the obtained values of $\omega_r$ are within these two bands, justifying the usage of the fluid dispersion relation.

The general resonance condition (Equation 4) can be simplified under weak relativistic approximation as
\begin{equation}
u_\perp^{2}/c^2+\left(u_\parallel/c-u_0/c\right)^2=r^2, r^2=N^2\cos^2\theta+2\left(\frac{n\Omega_{ce}}{\omega}-1\right)
\end{equation}
where $u_0/c=N\cos\theta$ and $N=kc/\omega_r$. Thus, the resonance curve in the $u$ space is a circle with the radius given by $r$ and the location of the center given by $u_0$. The growth rate is determined by both the details of the distribution function and the resonance curve. The instability grows when the resonance curve passes through regions with large and positive $\partial f/\partial u_\perp$. Such examples are shown in Figure 1 as black arcs or half circles. On the other hand, if the resonance curve passes through regions of small and/or negative values of $\partial f/\partial u_\perp$, the growth rate is either small or negative corresponding to wave damping. Such examples are plotted as yellow and green half circles in Figure 1. In our calculations, the absorption or damping effect of thermal electrons is taken into account.

\section{Parameter study on the Z-mode instability}
In this section, we present the parameter study on Z-mode excitation, focusing on the effect of \Tb and $v_e$. It is done by numerically solving the dispersion relation using the fluid approximation of warm plasmas to determine the mode frequency ($\omega_r$, normalized by $\Omega_{ce}$, Equation A.1), and integrating the kinetic dispersion relation to get the growth rate ($\gamma$, normalized by $\Omega_{ce}n_e/n_0$, Equations A.4). As mentioned, we limit our study in the parameter regime of $10 \leq \omega_{pe}/\Omega_{ce} \leq 30$, $0\leq T_0 \leq 8$ MK, and 0.15 c $\leq v_e \leq$ 0.4 c, with the weakly relativistic approximation. For each set of parameters, we calculate $\gamma$ within an appropriate range of ($\omega_r, \theta$). This yields a map of $\gamma$ over ($\omega_r, \theta$), through which we find the maximum growth rate $\gamma_{max}$ and the corresponding wave frequencies $\omega_r^{max}$ and $\theta_{max}$ for further analysis.

To demonstrate the method, we first present the study with a fixed $\omega_{pe}/\Omega_{ce}$ (= 15). This allows us to look into the individual contribution from various harmonics ($n$) and reach some general conclusions regarding the effect of $T_0$ and $v_e$. Then, we examine more details of their effect on the Z-mode instability over the given regime of $\omega_{pe}/\Omega_{ce}$.

\subsection{Effect of \Tb and $v_e$ on Z-mode growth with \omegarat = 15}
In Figures 3--5, we plot the map of the growth rate at various individual harmonic ($n$) (panels a--e) as well as their sum (panel f), for \omegarat = 15, $v_e = 0.15 \ c$ and $T_0 = $ 0, 2, 4 MK. From these maps, it is easy to tell the maximum growth rate for each harmonic. As expected, for different combination of parameters there always exists a specific harmonic $n_d$ at which the growth rate dominates over other harmonics, and only the two nearby harmonics with $n=n_d\pm1$ contribute significantly to the wave growth rate (given by the sum over harmonics). In addition, for all cases considered in this study, the Z-mode achieves the maximum growth rate always at the perpendicular or quasi-perpendicular direction. These features are consistent with earlier studies (e.g., \citealp{Winglee86}) that have used the DGH distribution functions for energetic electrons.

For $T_0 = 0$, i.e., the cold-plasma case, we have $n_d$ = 15 with $\gamma^n_{max} = 4.372$, $\omega^n_r = 15.032$, and $\theta^n = 88.8^\circ$, and similar values are obtained for the summed growth rate, indicating the dominance of this harmonic. For $T_0 = 2$ MK, we have $n_d$ = 16 with $\gamma^n_{max} = 4.303$, $\omega^n_r = 15.892$, and $\theta^n = 90^\circ$, and the summed growth rate also has similar or identical values. For $T_0 = 4$ MK, we have $n_d$ = 16 with $\gamma^n_{max} = 5.527$, $\omega^n_r = 15.761$, and $\theta^n = 90^\circ$, again the summed grow rate has similar or identical values. These results indicate that for warm plasmas the number of the dominant harmonic can deviate away from the value given by \omegarat (= 15).

This presents one important difference between the cold- and warm- plasma situations. For cold plasmas the frequency is fixed to the upper hybrid frequency $\omega_{UH}$, while for warm plasmas the Z-mode can grow in a much broader range covering the whole hybrid band and the bands higher (see Figure 2).

As seen from Figure 4, we see that the wave growth rates at $n=17$ and $n=15$ have smaller yet comparable maximum growth rates with that at $n_d=16$. The contributions from these three harmonics have been labeled in the map of the total wave growth rate (Figure 4f). From Figure 5, a very similar situation is observed for the three harmonics (15, 16 ($n_d$), and 17), yet the contribution of $n=15$ cannot be recognized from the summed map (Figure 5f). This is simply due to the result that the strong damping effect at $n_d = 16$ cancels the wave growth at $n=15$ within the corresponding regime of ($\omega_r, \theta$).

Another interesting observation is that, for both $T_0 = 2$ and 4 MK, the wave growth pattern with $\gamma > 0$ at the dominant harmonic splits into two parts by a strong absorption (or damping) regime, due to the presence of warm plasmas. In general, the thermal damping extends to a larger parameter space of ($\omega_r, \theta$) according to Figures 4 and 5.

The above results demonstrate the significance of including warm-plasma effect into the calculation of the Z-mode dispersion relation and their growth rate.

To further understand the resonance condition which decides the instability, we select three points in the ($\omega_r, \theta$) space, including the maximum growth rate at the dominant harmonic $n_d$, a weaker growth rate located nearby (at $\theta=88^\circ$), and a point in the strong wave damping region (see vertical arrows in panels c of Figures 3--5). The resonance curves given by corresponding parameters have been plotted onto the relevant velocity distribution functions, as shown in Figure 1.

Note that for cold plasmas, the absorption effect is due to the negative gradient of the distribution of energetic electrons $f_e$, and only two points, corresponding to the maximum and a weaker growth rate have been selected. As seen from Figure 1a, both curves pass through a significant part of the region with a positive gradient of $f_e$.

For warm plasmas, the resonance circle at the maximum growth rate is given by a zero of a very small value of $u_0 \ (\approx 0)$, corresponding to a nearly perpendicular propagation ($\theta \approx 90$). This makes the curve to sample the positive gradient region of $f$ most efficiently and thus leads to the maximum growth rate at the perpendicular propagation. The resonance curve corresponding to the weaker growth rate has a much larger radius and samples a part of the Maxwellian region with a significant negative gradient due to large number of electrons there, this makes the growth less efficient. On the other hand, the curve corresponding to strong damping passes through a significant section, including the central part, of the Maxwellian. This makes the growth rate negative and the Z-mode wave can not grow.

In Figure 6, we show the maps of the summed growth rate for $v_e = 0.15$ c (upper), 0.2 c (middle), and 0.3 c (lower) and $T_0 = 0$ (left), 2 (middle), 4 (right) MK. For ease of comparison, we re-present the results for $v_e = 0.15$ c. The variation trend is very clear from cold to warm plasmas with increasing $T_0$ and $v_e$, which can be summarized as: (1) the frequency range of wave growth increases significantly due to the warm-plasma effect on the wave dispersion relation, as already mentioned; (2) appearance of strong wave absorption or damping due to the inclusion of thermal electrons, in particular, the growth region at the dominant harmonic ($n_d$) splits into two parts by the damping of Maxwellian; (3) the maximum wave growth always appears at perpendicular or quasi-perpendicular direction, as required by the resonance curve to sample the most positive gradient of $f$, as elaborated above; (4) with increasing $v_e$, a clear trend with non-negligible contributions from more harmonics and a slight decrease of $\gamma_{max}$ are observed.


The above studies are based on a few discrete values of $T_0$ and $v_e$, as done in most earlier studies. Here in Figure 7 we show variation profiles of the three parameters of Z-mode instability ($\gamma_{max}, \omega^r_{max}, \theta_{max}$) with $T_0$ and $v_e$. This allows us to explore more details of the parameter dependence.

As seen from the left panels of Figure 7, for $v_c \leq 0.25$ c the maximum growth rate manifests an obvious oscillation with increasing $T_0$. For instance, for $v_e = 0.15$ c, $\gamma_{max}$ reaches its peak of 5.349 at 3 MK; for $v_e = 0.2$ c, $\gamma_{max}$ reaches its peak of 2.926 at 4.5 MK. On the other hand, for $v_e > 0.25$ c, the oscillation pattern is not significant, in other words, $\gamma_{max}$ is almost independent of $T_0$. The wave frequencies $\omega^r_{max}$ vary within the hybrid band for lower $T_0$ and may jump into the band higher for larger $T_0$. The jumping point is different for different $v_e$. For $v_e = 0.15$ c, $\omega^r_{max}$ jumps from 15.270 to 15.932 around 1.125 MK, and for $v_e = 0.25$ c, $\omega^r_{max}$ jumps from 15.383 to 15.776 around 3.375 MK. Before the jumps, $\omega^r_{max}$ increases gradually while after the jumps $\omega^r_{max}$ presents a slow declining trend. This gives the interesting behavior of the stepwise variation of $\omega^r_{max}$. The stepwise jumping point happens at different values of $T_0$ for different $v_e$. This behavior is mainly due to the increase of the number of the dominant harmonic ($n_d$) by unity, in response to the continuous increase of $T_0$. Note that the values of $n_d$ have been written onto the upper panel. Further discussion will be presented in the last section.

For large $T_0$ and low $v_e$, it can be seen that $\omega^r_{max}$ is in the band higher than the hybrid band. We highlight that $\omega^r_{max}$ can appear in both bands, its exact values depend on the values of $T_0$ and $v_e$. These results are significant to studies on t-IV solar radio bursts with or without zebra patterns since this basically determines the frequencies of emission. It also affects any further studies to infer the magnitude of the magnetic field strength and plasma density on the basis of radio spectral data. More discussion is presented in the last section.

The propagation angle at the maximum growth rate ($\theta_{max}$) is or very close to 90$^\circ$, consistent with earlier studies on Z-mode instability using the DGH distribution of energetic electrons. Here, we show that $\theta_{max}$ varies in accordance with $\omega^r_{max}$. This is because that the two parameters must vary coherently to meet the resonance condition (Equation 4).

From the right panels of Figure 7, we see that both $\gamma_{max}$ and $\omega^r_{max}$ decline in-general with increasing $v_e$. This trend becomes not significant when $v_e$ is large enough, say, $v_e > 0.25$ c. In other words, when $v_e$ is large enough, both wave parameters only weakly depend on $v_e$. Again, a stepwise variation of $\omega^r_{max}$ exists when $v_e$ increases continuously. This happens at $v_e$ = 0.189 c for $T_0 = 2$ MK, and at $v_e$ = 0.255 c for $T_0 = 4$ MK, also due to the jump of the dominant harmonic number by unity. Note that the number $n_d$ has been written onto the upper panel. The wave grows and propagates mainly along the perpendicular or quasi-perpendicular direction, and the angle $\theta_{max}$ presents a variation pattern in accordance with that of the $\omega^r_{max}$.


\subsection{Effect of \Tb and \vc on Z-mode growth with 10 $\le$ \omegarat $\le$ 30}

In this subsection, we investigate the effect of \omegarat on $\gamma_{max}$ of the instability, as done in many earlier studies. Varying \omegarat can be understood as varying the magnitude of the background magnetic field and the plasma density. Thus, this makes a preliminary study of wave excitation in inhomogeneous media if assuming the spatial scale of the inhomogeneity is much larger than the wavelength.

The ratio \omegarat varies in a range of [10, 30]. For any specific value of $\omega_{pe}/\Omega_{ce}$, we conduct the parameter study as those described above and find the corresponding $\gamma_{max}$, then plot the profiles of $\gamma_{max}$ versus $\omega_{pe}/\Omega_{ce}$. A major purpose of this subsection is to clarify some inconsistent results given by earlier publications, as stated earlier.

From Figures 8a--8b, we plot such profiles for different values of $T_0$ while fixing $v_e$ at 0.2 c. The most prominent feature of the profiles is their oscillations, with quasi-periodic peaks and valleys. The distance between neighboring peaks is about one $\Omega_{ce}$. This pattern is well known and has been applied to explain the presence of zebra patterns of solar t-IV radio bursts (e.g., \citealp{Winglee86,Yasnov04,Kuznetsov07,Benacek17}). Another important feature is the overall increasing trend of $\gamma_{max}$ with increasing $\omega_{pe}/\Omega_{ce}$ and the very-weak-yet-discernible decreasing trend with increasing $T_0$. This result is different from that presented by \cite{Benacek17} who show that $\gamma_{max}$ decreases in-general with increasing $\omega_{pe}/\Omega_{ce}$. The difference might be due to the specific simplifications used in their model. For example, they have neglected the term containing the parallel gradient of $f_e$ (i.e., $\partial f_e/\partial u_\parallel$) when calculating the growth rate. In addition, the specific peaks shift towards lower values of \omegarat with increasing $T_0$. This is basically consistent with those given by \cite{Benacek17} (see Figure 6 of this reference).

The ratio between values of $\gamma_{max}$ at neighboring peaks and bottoms can be used to characterize the flatness of the profile, as plotted in Figure 8c. It can be seen that the ratio declines with increasing $\omega_{pe}/\Omega_{ce}$. For instance, for \omegarat = 15, the ratio is about 2.1 at $T_0 = 1$ MK and $\sim$ 1.9 at $T_0 = 3$ MK, while for \omegarat = 25, the ratio is $\sim$ 1.6 at $T_0 = 1$ MK and $\sim$ 1.3 at $T_0 = 3$ MK. The flatness variation trend of $\gamma_{max}$ versus $\omega_{pe}/\Omega_{ce}$ is basically consistent with the result of \cite{Benacek17}. For $\omega_{pe}/\Omega_{ce} < 15$, the peak-bottom ratio decreases with increasing $T_0$, while the variation of flatness of $\gamma_{max}$ with \Tb is not very regular for larger values of $\omega_{pe}/\Omega_{ce} \ (> 15)$. The latter might be caused by inregular oscillations and steep changes of $\gamma_{max}$ at the bottom of the profiles.

In Figure 9, we show similar profiles of $\gamma_{max}$ with \omegarat for different $v_e$ while fixing $T_0$ to be 2 MK. A very weak overall increasing trend of $\gamma_{max}$ with increasing \omegarat can be identified from the upper panel. From the profiles of peak-bottom ratio (Figure 9b), it can be seen that the ratio declines with increasing \omegarat and also with increasing $v_e$. For $v_e \geq 0.3$ c and \omegarat $\geq 15$, the ratio is around or less than 1.2, indicating that the zebra pattern may not be recognizable under these conditions. The result presented in Figure 9 is basically consistent with those presented by \cite{Winglee86} and \cite{Benacek17}. Here we extend the calculations to a larger parameter regime of \omegarat.

\section{Summary and discussion}

Many earlier studies have studied the Z-mode instability to understand the origin of zebras of t-IV bursts. Among various parameters, the ratio of $\omega_{pe}$ and $\Omega_{ce}$ (decided by the magnetic field strength and plasma density) plays a major role on the maximum of the instability growth rate ($\gamma_{max}$). In addition to this ratio, the temperature of background plasma ($T_0$) and the energy of energetic electrons ($v_e$) are also important. Earlier studies only considered a few discrete values of the two parameters and revealed inconsistent results. Here we revisit the problem with a more complete parameter study on the effect of these parameters. The parameter regimes of interest, relevant to latest observations on t-IV sources, are taken to be
$10 \leq \omega_{pe}/\Omega_{ce} \leq 30$, $0\leq T_0 \leq 8$ MK, and 0.15 c $\leq v_e \leq $ 0.4 c.

For a specific value of \omegarat (= 15), it was found that $\gamma_{max}$ presents a general decreasing trend with increasing $v_e$, and an obvious oscillation with increasing $T_0$  for $v_e = 0.15 - 0.25$ c while it is almost independent of $T_0$ for larger $v_e$. In addition, with increasing $T_0$ and $v_e$, the frequency at $\gamma_{max}$ presents step-wise profiles with jumps separated by gradual or weak variations in the range covering the hybrid band and the band higher. The propagation angle $\theta_{max}$ varies accordingly as constrained by the resonance condition. The cause of these jumps, as mentioned in Section 3, is due to the change of the dominant harmonic number ($n_d$) by unity. Further explanation is given as follows. To achieve the maximum growth rate, the resonance curve must sample the appropriate region of the velocity distribution ($f$), i.e., with a large positive velocity gradient of $f$ and enough number of particles along the curve. This puts strong constraints on $\omega_r \ (k)$ and $\theta$, at which the maximum growth rate is obtained. These parameters, along with $n$, decide the resonance curve. In Figure 7, we increase $T_0$ and $v_e$ gradually. The two parameters change the velocity gradient of $f$ and thus the resonance condition at $\gamma_{max}$. Generally speaking, a larger $T_0$, i.e., a more-expanded Maxwellian distribution, corresponds to a more-expanded resonance curve at $\gamma_{max}$, while $f$ with a larger $v_e$ will result in the opposite trend of the resonance curve at $\gamma_{max}$. In addition, the dispersion relation of Z-mode changes with $T_0$. This further affects the growth rate.

As seen from our results, below certain thresholds, the maximum rate can still be obtained for fixed harmonic number $n$ with gradually-changing values of $\omega_r$, $\gamma_L$, and $\theta$. Yet, above the thresholds, the maximum growth rate moves to the nearby harmonic, $n+1$ for increasing $T_0$ and $n-1$ for increasing $v_e$, due to the above-mentioned opposite trend of $T_0$ and $v_e$ on the velocity distribution function $f$. This change of harmonic number leads to the jumps of various parameters observed in Figure 7.

One explanation of zebras is that each stripe is given by a peak of growth rate which appear for continuous variation of $\omega_{pe}/\Omega_{ce}$. For cold plasmas the peak is reached when $\omega_{UH} = s \Omega_{ce}$, i.e., when the upper hybrid frequency equals to a harmonic of electron cyclotron frequency. This is not correct for warm plasmas. Taking the warm-plasma effect into account, within a range of $10 \le $ \omegarat $\le 30$ we studied the influences of $T_0$ and $v_e$ on the variation of the peaks and relevant maximum-minimum ratios of growth rate. It was found that the ratios always decline and the location of peaks shift towards lower values of $\omega_{pe}/\Omega_{ce}$, with increasing $\omega_{pe}/\Omega_{ce}$. The ratios do not present a simple variation trend with increasing $T_0$, and it in-general declines with increasing $v_e$ for $v_e \leq 0.3$ c. For larger $v_e$, the ratio remains around or less than 1.2. Such small values of ratio may lead to continuum without recognizable zebras.

During solar flares, both heating and particle acceleration take place. This leads to change of the plasma temperature ($T_0$) and energy of energetic electrons ($v_e$). Thus, it is natural to suggest that the frequency variations of zebra stripes may be partially due to the on-going heating and acceleration processes (e.g., \citealp{Yasnov04}). For example, according to our calculations, with continuous plasma heating and particle acceleration, the wave frequency may change either gradually or suddenly. And the maximum growth rate also changes accordingly. This may affect the morphology of the stripes and their emission intensity. Thus, the calculations presented here are significant to explanations of zebras and further studies to infer coronal parameters, such as the magnetic field strength in the source region (e.g., \citealp{Tan12}). In addition, it should be highlighted that both magnetic field and plasma density change rapidly during solar flares, and this may also have an important effect on the wave growth rate as well as the presence of zebra stripes and their spectral morphology. Further studies to explore origins of various types of zebras should take these factors into account.

\section*{Acknowledgements}
This study is supported by the National Natural Science Foundation of China (11790303 (11790300), 11750110424 and 11873036). X.K. also acknowledges the support from the Young Elite Scientists Sponsorship Program by China Association for Science and Technology, and the Young Scholars Program
of Shandong University, Weihai. The authors are grateful to the anonymous referee for valuable comments.

\appendix

\section{The growth rate and dispersion tensor of Z-mode instability for warm plasmas}

The wave frequency in a collisionless Vlasov-Maxwell system is written as $\omega = \omega_r + i\gamma$, where $\omega_r$ is determined by the dispersion relation using the following fluid approximation of warm plasmas for X (Z) mode
\begin{equation}
\text{Re}\overleftrightarrow{\Lambda}(\vec{k},\omega_r)=
\left( \begin{array}{ccc}
 -N^2\cos^2\theta & 0 & N^2\sin\cos\theta \\
 0 & -N^2 & 0 \\
 N^2\sin\cos\theta & 0 & -N^2\sin^2\theta
\end{array}
\right )+\overleftrightarrow{\epsilon}=0,
\end{equation}
\begin{equation}
\overleftrightarrow{\epsilon}=\overleftrightarrow{I}-\frac{\omega_{pe}^2}{\omega_r^2}\frac{\overleftrightarrow{C}_e}{\Delta_e}, \Delta_e=(1-\frac{\Omega_{ce}^2}{\omega_r})(1-3N^2v_0^2\cos^2\theta)-3N^2v_0^2\sin^2\theta,
\end{equation}
\begin{equation}
\overleftrightarrow{C}_e=
\left( \begin{array}{ccc}
 1-3N^2v_0^2\cos^2\theta & -i\frac{\Omega_{ce}}{\omega_r}(1-3N^2v_0^2\cos^2\theta) & 3N^2v_0^2\sin\theta\cos\theta \\
 i\frac{\Omega_{ce}}{\omega_r}(1-3N^2v_0^2\cos^2\theta) & 1-3N^2v_0^2 & i\frac{\Omega_{ce}}{\omega_r}3N^2v_0^2\sin\theta\cos\theta \\
 3N^2v_0^2\sin\theta\cos\theta & -i\frac{\Omega_{ce}}{\omega_r}3N^2v_0^2\sin\theta\cos\theta & 1-\frac{\Omega_{ce}^2}{\omega_r^2}-3N^2v_0^2\sin^2\theta
\end{array}
\right ),
\end{equation}
 where $N=kc/\omega_r$ and $k$ are the refractive index and the wave number, respectively, and $\theta$ is the angle of propagation (i.e., the angle between $\vec{k}$ and $\vec{B}$), and $v_0=\sqrt{k_BT_0/m_e}$ is the thermal velocity of background electrons.

Under the assumption of $\omega_r \gg \left|\gamma\right|$, the growth rate $\gamma$ is given by
\begin{equation}
\gamma=-\frac{\text{Im}\Lambda(\vec{k},\omega_r)}{\frac{\partial}{\partial\omega_r}\text{Re}\Lambda(\vec{k},\omega_r)}.
\end{equation}
$\text{Im}\Lambda(\vec{k},\omega_r)$ is the imaginary part of the kinetic dispersion relation given by \citep[e.g.][]{Baldwin69}
\begin{equation}
\begin{split}
\text{Im}\Lambda(\vec{k},\omega_r)=&2\pi\frac{\omega_{pe}^2}{\omega_r^2}\int_{-\infty}^{+\infty}du_{\parallel}\int_{0}^{+\infty}du_{\perp}\Bigg\{\frac{u_{\parallel}}{\gamma_L}\left(u_\perp\frac{\partial}{\partial u_\parallel}-u_\parallel\frac{\partial}{\partial u_\perp}\right)\times f(u_\perp,u_\parallel)\hat{e}_z\hat{e}_z+\\
&\omega_r\left[\frac{\partial}{\partial u_\perp}+\frac{k_\parallel}{\gamma_L\omega_r}\left(u_\perp\frac{\partial}{\partial u_\parallel}-u_\parallel\frac{\partial}{\partial u_\perp}\right)\right]\times f(u_\perp,u_\parallel)\sum_{n=-\infty}^{\infty}\frac{\overleftrightarrow{T_n}(b)}{\gamma_L\omega_r-n\Omega_{ce}-k_\parallel u_\parallel}\Bigg\}
\end{split},
\end{equation}
\begin{equation}
\overleftrightarrow{T_n}(b)=
\left( \begin{array}{ccc}
\frac{n^2\Omega_{ce}^2}{k_\perp^2}J_n^2(b) & -i\frac{n\Omega_{ce}}{k_\perp}u_\perp J_n(b)J_n^\prime(b) & \frac{n\Omega_{ce}}{k_\perp}u_\parallel J_n^2(b)\\
i\frac{n\Omega_{ce}}{k_\perp}u_\perp J_n(b)J_n^\prime(b) & u_\perp^2J_n^{\prime2}(b) & iu_\perp u_\parallel J_n(b)J_n^\prime(b)\\
\frac{n\Omega_{ce}}{k_\perp}u_\parallel J_n^2(b) & -iu_\perp u_\parallel J_n(b)J_n^\prime(b) & u_\parallel^2J_n^2(b)
\end{array}
\right ),
\end{equation}
where $u_\parallel=p_\parallel /m_e=\gamma_L v_\parallel$, $u_\perp=p_\perp /m_e=\gamma_L v_\perp$, $\omega_{pe}=\sqrt{n_ee^2 /m_e\varepsilon_0}$ is the plasma frequency, $\gamma_L=\left(1-\frac{v^2}{c^2}\right)^{-1/2}$ is the Lorentz factor, $f(u_\perp,u_\parallel)$ is the total distribution function of electrons, and $J_n(b)$ the first-kind Bessel function of the \emph{n}th order, $J_n^\prime(b)$ is its partial derivative with respect to $b \ (=k_\perp u_\perp /\Omega_{ce})$.


 \begin{figure*}
   \centerline{\includegraphics[trim= 8.0cm -1cm -2.5cm 1cm,scale=0.75,angle=90]{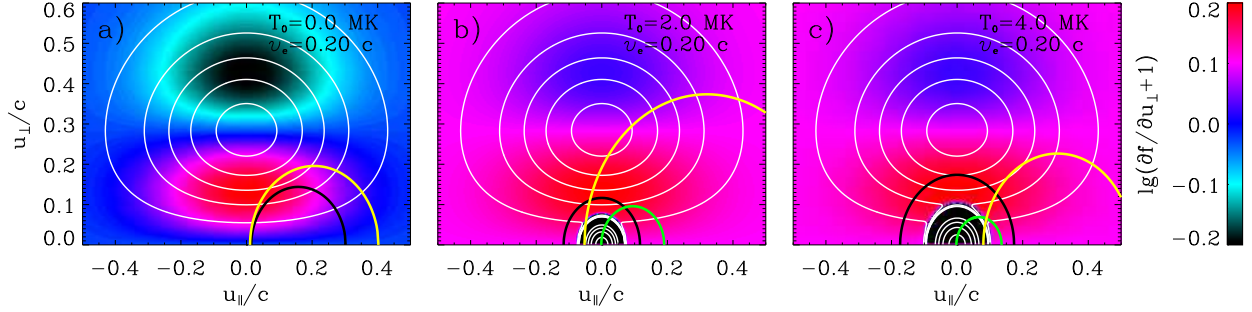}
              }
              \caption{Three examples of the total electron distribution function (represented by white contours), superposed onto the map of $lg(\partial f/\partial u_\perp+1)$ for different values of \Tb(0, 2, 4 MK) and a fixed value of \vc= 0.2 \ccc. The black arcs or half circles represent resonance curves associated with the maximum growth rate. Yellow circles are given by parameters yielding smaller growth rate and green ones are associated with parameters of strong wave damping region. The parameters can be seen from Figures 3--5, pointed at with arrows of corresponding colors.
              }
   \label{f1}
   \end{figure*}
\begin{figure*}
   \centerline{\includegraphics[trim= 0.5cm 2.2cm 8cm -4cm,scale=0.65,angle=90]{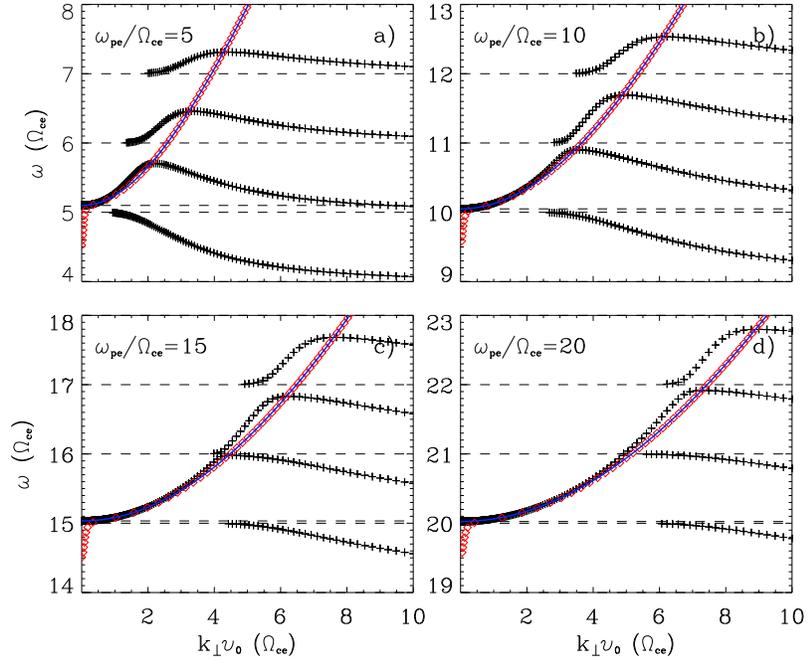}
              }
              \caption{Dispersion relations given by the fluid (red circles) and the kinetic theory (black plus signs) for Z-mode waves in warm plasmas of Maxwellian distribution, for \omegarat = 5, 10, 15, and 20. The blue lines correspond to the dispersion curves of the upper hybrid waves ($\omega^2=\omega^2_p+\Omega^2_{ce}+3k^2_\perp v^2_0$).
              }
   \label{f2}
   \end{figure*}
\begin{figure*}
   \centerline{\includegraphics[trim= 3.5cm 0.5cm -0.5cm 1cm,scale=0.70,angle=90]{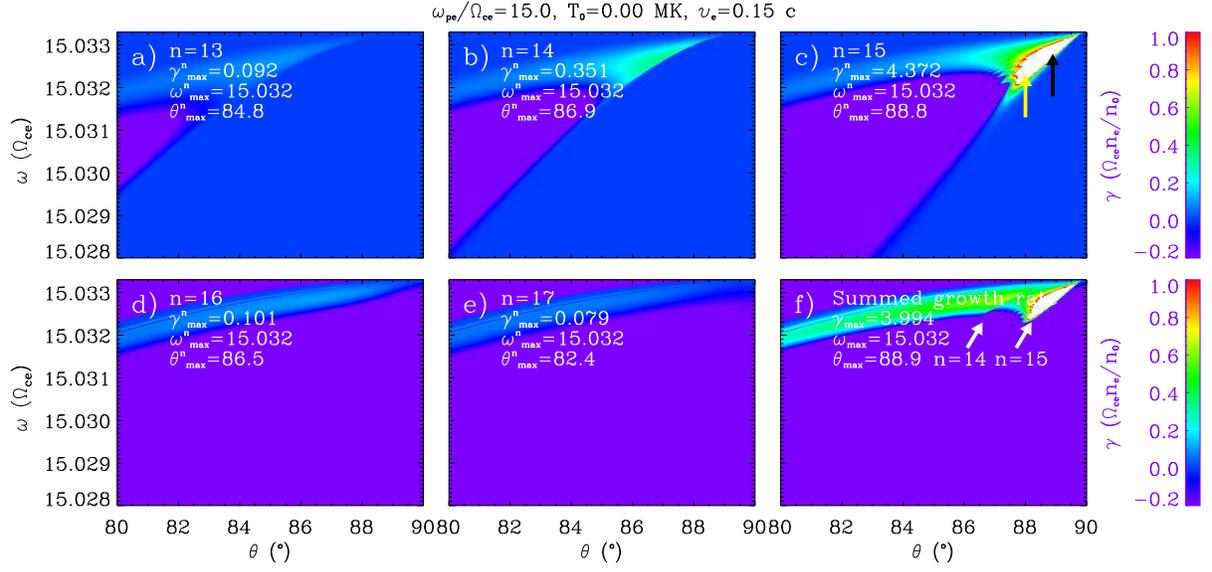}
              }
              \caption{The maps of growth rate at various harmonics ($n$) (panels a--e) as well as their sum (panel f) for \omegarat = 15, $v_e = 0.15 \ c$ and $T_0 = $ 0 MK. Two arrows in panel c point to the location with the maximum growth rate ($\gamma_{max}$, in black) and a nearby point with a slightly weaker growth rate (at $\theta = 88^\circ$, in green). The corresponding resonance curves given by these points with same color have been plotted onto the maps of relevant velocity distribution functions, as shown in Figure 1a.
              }
   \label{f3}
   \end{figure*}
\begin{figure*}
   \centerline{\includegraphics[trim= 3.5cm 0.5cm 5cm 1cm ,scale=0.70,angle=90]{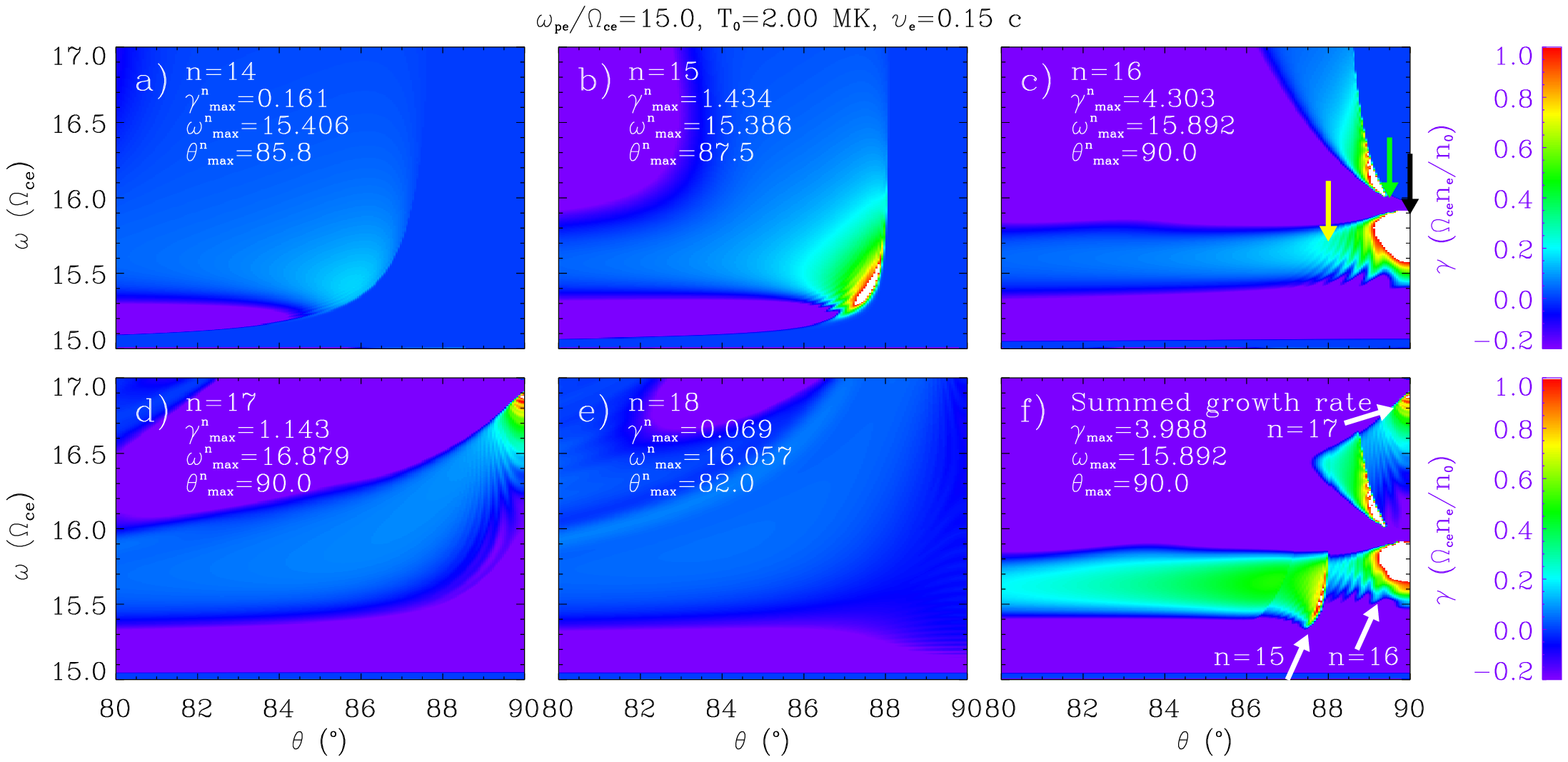}
              }
              \caption{The maps of growth rate at various harmonics ($n$) (panels a--e) as well as their sum (panel f) for \omegarat = 15, $v_e = 0.15 \ c$ and $T_0 = $ 2 MK. Three arrows in panel c point to the location with the maximum growth rate ($\gamma_{max}$, in black), a nearby point with a slightly weaker growth rate (at $\theta = 88^\circ$, in yellow) and a point in the strong wave damping region ($\theta=89.5^\circ, \ \omega_r=16$, in green). The corresponding resonance curves given by these points with same color have been plotted onto the maps of relevant velocity distribution functions, as shown in Figure 1b.
              }
   \label{f4}
   \end{figure*}
\begin{figure*}[!ht]    
   \centerline{\includegraphics[trim= 3.5cm 0.5cm -0.5cm 1cm ,scale=0.70,angle=90]{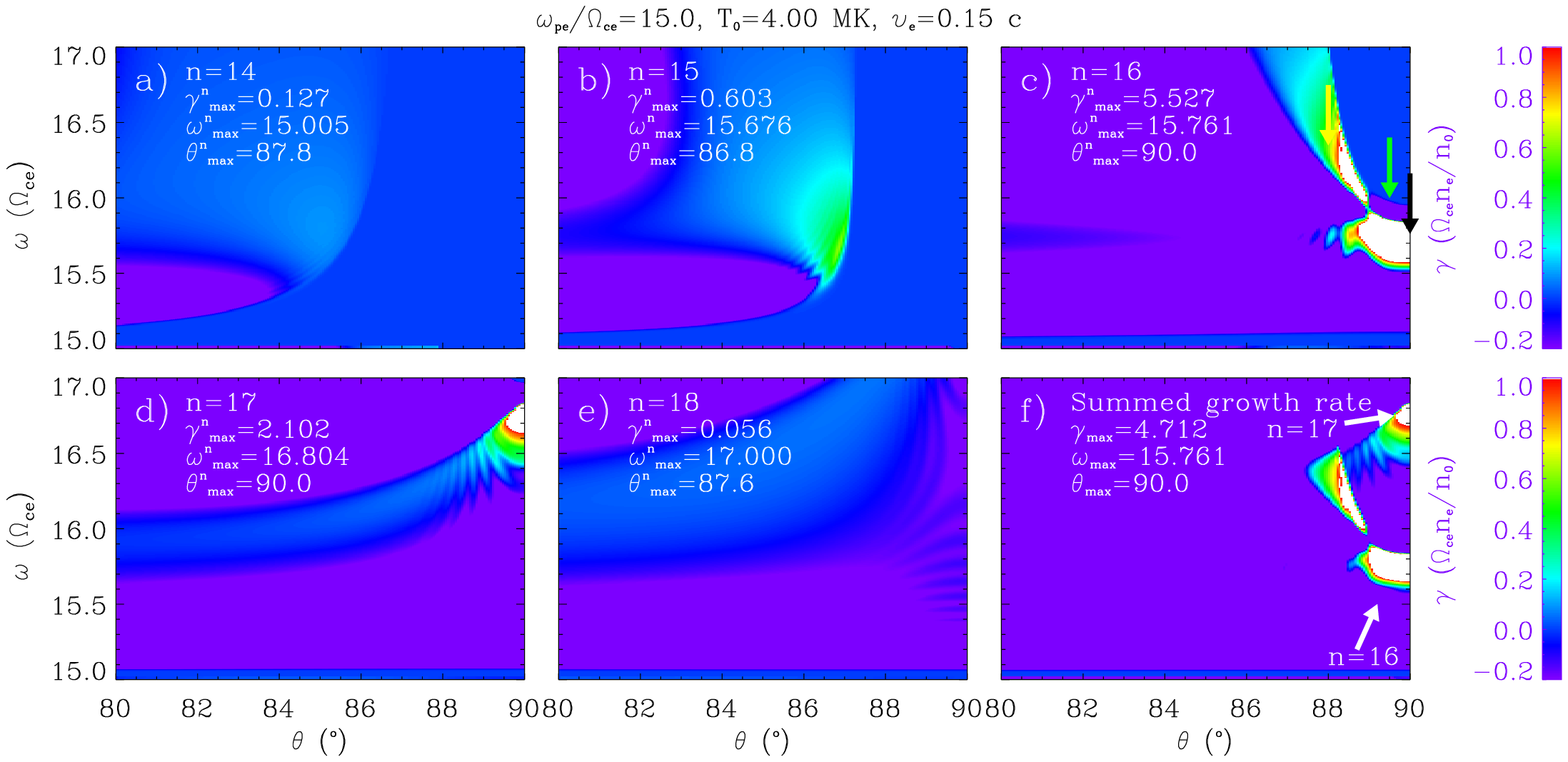}
              }
              \caption{The maps of growth rate at various harmonics ($n$) (panels a--e) as well as their sum (panel f), for \omegarat = 15, $v_e = 0.15 \ c$ and $T_0 = $ 4 MK. Three arrows in panel c point to the location with the maximum growth rate ($\gamma_{max}$, in black), a nearby point with a slightly weaker growth rate (at $\theta = 88^\circ$, in yellow) and a point in the strong wave damping region ($\theta=89.5^\circ, \ \omega_r=16$, in green). The corresponding resonance curves given by these points with same color have been plotted onto the maps of relevant velocity distribution functions, as shown in Figure 1c.
              }
   \label{f5}
   \end{figure*}
\begin{figure*}[!ht]    
   \centerline{\includegraphics[trim= 0.5cm 2.5cm 2cm 1cm ,scale=0.70,angle=90]{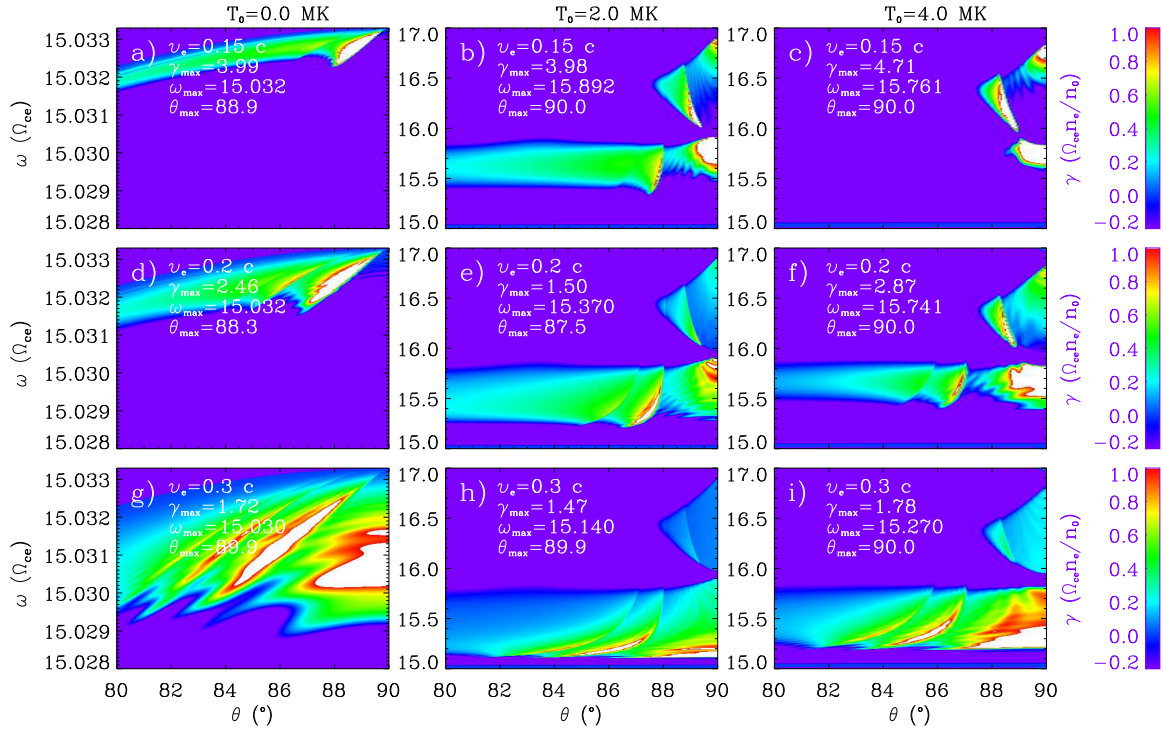}
              }
              \caption{Maps of the summed growth rate for $v_e$ = 0.15 c (upper), 0.2 c (middle), and 0.3 c (lower) and $T_0$ = 0 (left), 2 (middle), 4 (right) MK.}
   \label{f6}
   \end{figure*}
\begin{figure*}[!ht]    
   \centerline{\includegraphics[trim= -1cm 2.2cm -4.5cm 1cm,scale=0.65,angle=90]{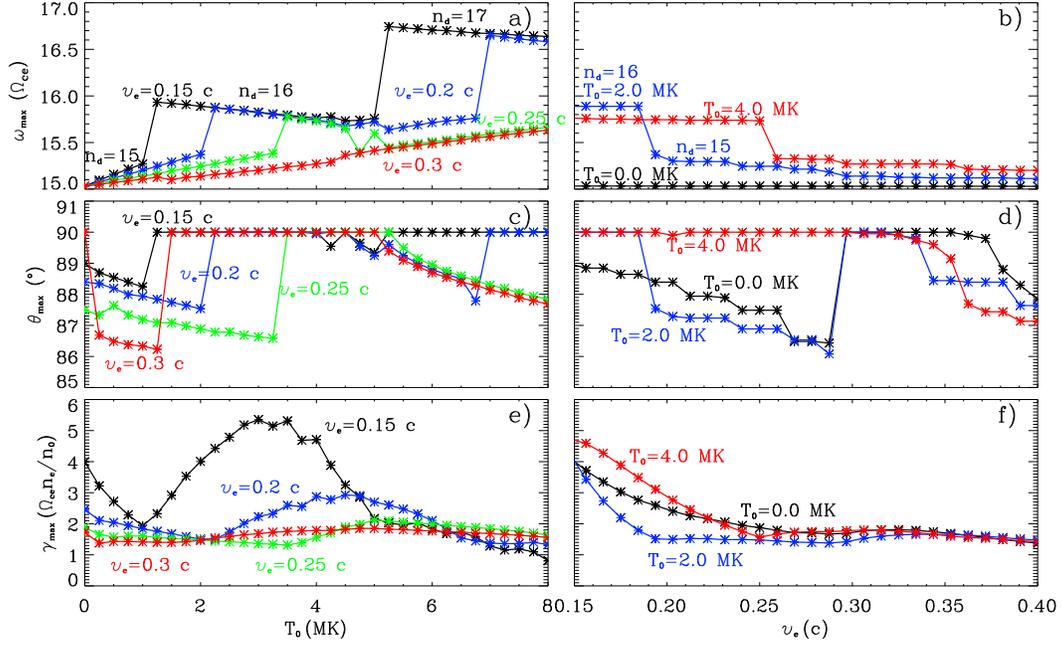}
              }
              \caption{Variation profiles of $\omega^r_{max}, \theta_{max}, \gamma_{max}$ with $T_0$ (left) and $v_e$ (right).
              }
   \label{f7}
   \end{figure*}

\begin{figure*}[!ht]    
   \centerline{\includegraphics[trim= 1.0cm 3cm 0.5cm 1cm,scale=0.75]{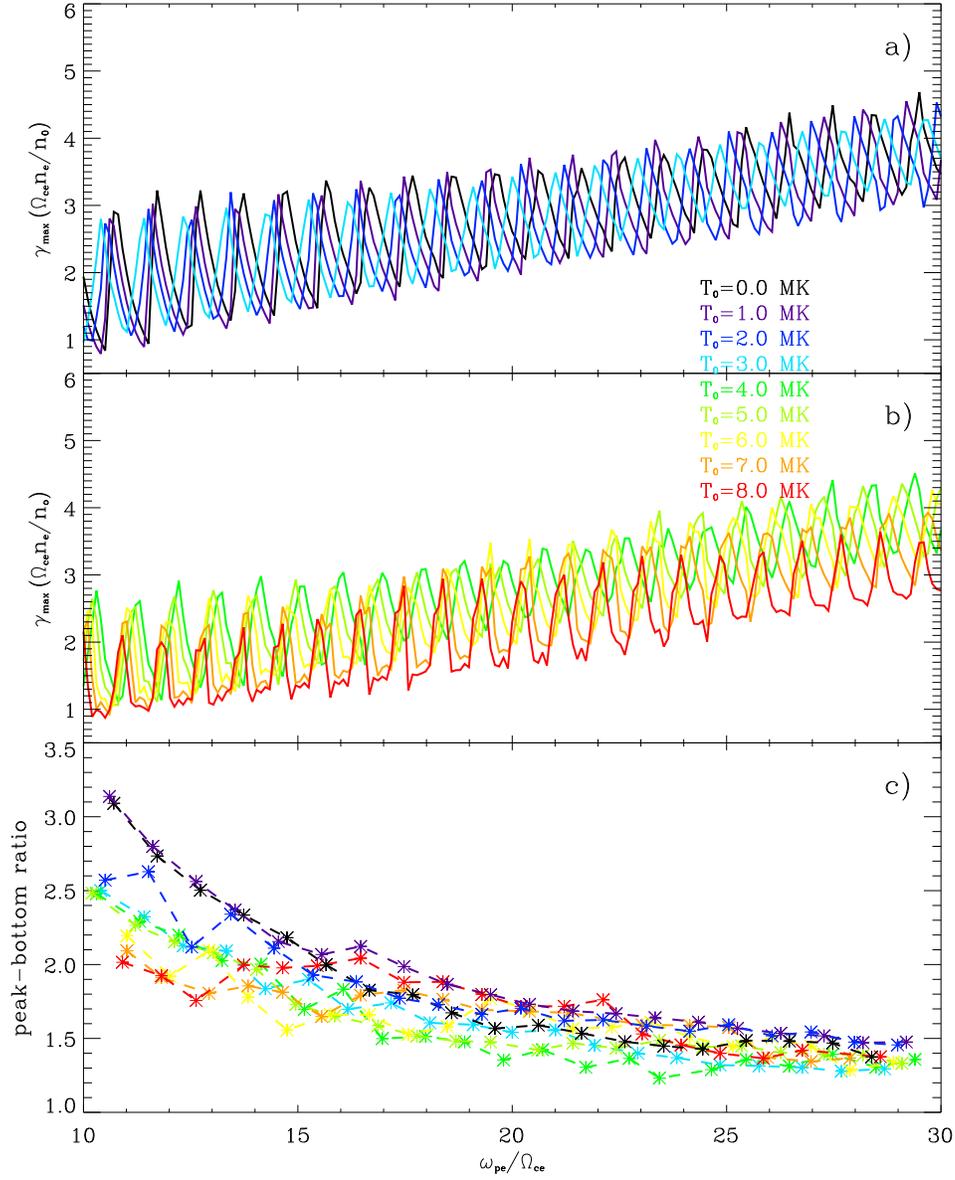}
              }
              \caption{(a--b) \gammamax as a function of \omegarat with \Tb = 0--8 MK and \vc = 0.2 c. Different colors represent different values of $T_0$. (c) Ratios of maxima and neighbouring minima of \gammamax versus $\omega_{pe}/\Omega_{ce}$, to indicate the flatness of the profiles shown in the upper panel.
              }
   \label{f7}
   \end{figure*}
\begin{figure*}[!ht]    
   \centerline{\includegraphics[trim= 1.0cm 10cm 0.5cm 0cm,scale=0.75]{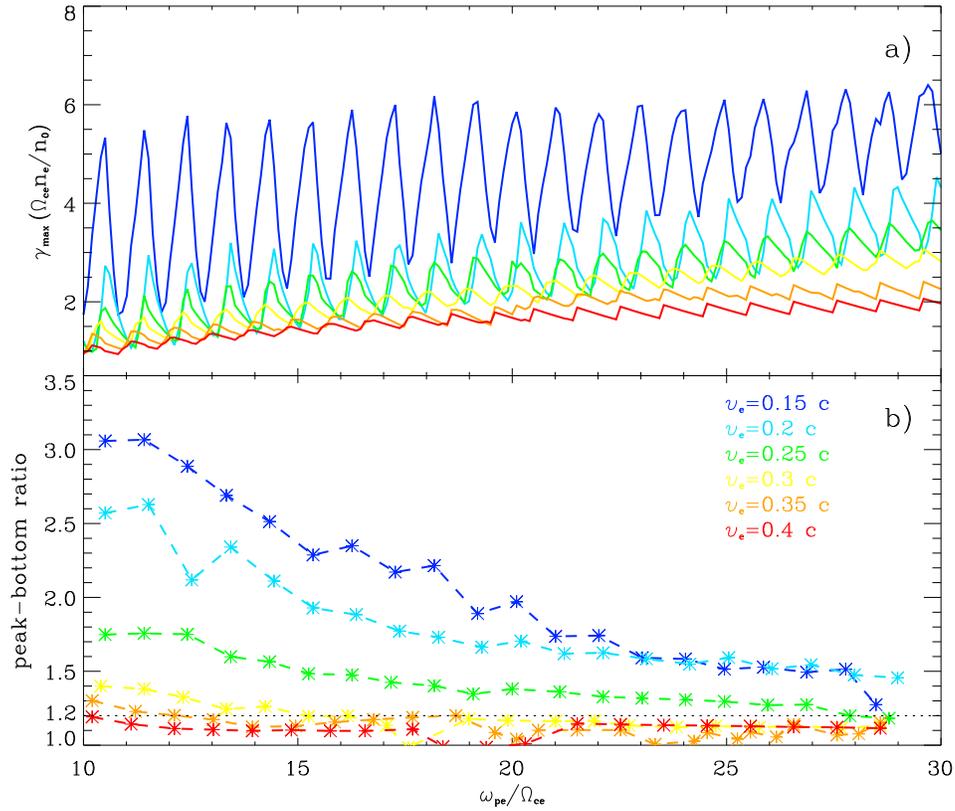}
              }
              \caption{(a) \gammamax as a function of \omegarat with \vc = 0.15--0.4 c and \Tb = 2 MK. Different colors represent different values of $v_e$. (b) The peak - bottom contrast versus $\omega_{pe}/\Omega_{ce}$, to indicate the flatness of the profiles shown in the upper panel. The black dashed line is at 1.2.
              }
   \label{f7}
   \end{figure*}
\end{document}